-------------------------------------------------------------------------------
\documentstyle[12pt]{report}
\begin{document}
\begin{centering}
\large\bf A New Young Diagrammatic Method For\\
Kronecker Products of O(n) and Sp(2m)
\vskip .6truecm
\normalsize Feng Pan$^{1\dagger}$, Shihai Dong$^{2}$ and J. P. Draayer$^{1}$
\vskip .2cm
\noindent{\small\it $^{1}$Department of Physics {\normalsize\&} Astronomy,
Louisiana State University,}\\
{\small\it Baton Rouge, LA 70803-4001}\\
\vskip .1cm
{\small\it $^{2}$Department of Physics, Liaoning Normal University,
Dalian 116029, P. R. China}\\
\vskip 1truecm
{\bf Abstract}\\
\end{centering}
\vskip .5cm
   \normalsize A new simple Young diagrammatic method for Kronecker products of
O(n) and Sp(2m) is proposed based on representation theory of Brauer algebras.
A general procedure for the decomposition of tensor products of representations
for O(n) and Sp(2m) is outlined, which is similar to that for U(n) known as
the Littlewood rules together with trace contractions from a Brauer algebra
and some modification rules given by King.
\vskip 2.5cm
\noindent {-----------------------------------------}\\
\noindent $^{\dagger}$On leave from Department of Physics,
Liaoning Normal Univ., Dalian 116029, P. R.~~China

\newpage

\begin{centering}
{\large 1. Introduction}\\
\end{centering}
\vskip .4truecm
   Representation theory of orthogonal and symplectic groups plays an important
role in many areas of physics and chemistry. It arises, for example, in the
description of symmetrized orbitals in quantum chemistry$^{[1]}$, fermion
many-body theory$^{[2]}$, grand unification theories for elementary
particles$^{[3]}$, supergravity$^{[4]}$, interacting boson and
fermion dynamical symmetry models for nuclei$^{[5-8]}$, nuclear symplectic
models$^{[9-10]}$, and so on.
\vskip .3cm
   Reductions of Kronecker products of representations of O(n) and Sp(2m)
groups
were outlined in a series of works of King and his
collaborators$^{[11-15]}$ based
on the pioneering work of Murnaghan$^{[16]}$, Littlewood$^{[17-18]}$, and 
Newell$^{[19]}$
on character theory and Schur functions. A similar approach was then revisited
by Koike and Terada$^{[20]}$, in which some main points were rigorously proved.
On the other hand, a Young diagrammatic method for Kronecker products for
Lie groups
of B, C, and D types were proposed by Fischer$^{[21]}$. However, as pointed
out by
Girardi et al$^{[22-23]}$, rules for the decomposition of tensor products
for SO(n)
and Sp(2m) given in [21] are numerous; and some of them are even ambiguous.
After
introducing generalized Young tableaux, with negative rows for describing
SO(2m),
Girardi et al gave a formula to compute the Kronecker products for SO(n)
and Sp(2m)
in [22-23]. The formula can be used to compute both tensor and spinor
representations
of SO(n) and Sp(2m). However, no proof was given for their formula.
Littelmann in [24] proposed another Young tableau method to compute
Kronecker product of
some simply connected algebraic groups based on the character theory. The feature
of the method is that it does not use the representation theory of symmetric
group. later, Nakashima  proposed a Crystal graph base [25] 
together with the generalized Young diagrams for the same problem. This method
applies equally well for the $q$-analogue of the universal enveloping algebras of
$A$, $B$, $C$, and $D$ types$^{[26]}$.
\vskip .3cm
   In addition to the usefulness of these groups in many applications, the
decomposition
of the Kronecker products of orthogonal and symplectic groups have long
been an interesting
problem in mathematics, which was first considered by Weyl$^{[27]}$ and
Brauer$^{[28]}$.
Besides the works of mentioned above, there are
many other similar works. For example, Berele discussed a similar problem
for the symplectic
case in [29], and Sundaram for the orthogonal case in [30].
\vskip .3cm
   In this paper, we will outline a new simple Young diagrammatic method
for the Kronecker
products of O(n) and Sp(2m). Our procedure is mainly based on the induced
representation of
the Brauer algebra $D_{f}(n)$, which applies to O(n) and Sp(2m) because of
the well-known
Brauer-Schur-Weyl duality relation between $D_{f}(n)$ and O(n) or Sp(2m).
This relation has
already enabled us to derive Clebsch-Gordan coefficients and Racah
coefficients of the quantum
group $U_{q}(n)$ from induction and subduction coefficients of Hecke
algebras$^{[31-32]}$,
and Racah coefficients of O(n) and Sp(2m) from subduction coefficients of Brauer algebra$^{[33]}$.
\vskip .3cm
   In Section 2, we will give a brief introduction to Brauer algebras.
Induced representations
of the Brauer algebra $S_{f_{1}}\times S_{f_{2}}\uparrow D_{f}(n)$
will be discussed in
Section 3, which are important for our purpose. In Section 4, we will
outline a new simple Young
diagrammatic method for the decomposition of the Kronecker products for
O(n) and Sp(2m). A
concluding remark will be given in Section 5.
\vskip .5cm
\begin{centering}
{\large 2. Brauer algebra $D_{f}(n)$}\\
\end{centering}
\vskip .3cm
   The Brauer algebra $D_{f}(n)$ is defined algebraically by $2f-2$ generators
$\{ g_{1},~g_{2},~\cdots,~g_{f-1},$\\
$e_{1},~e_{2},~\cdots,~e_{f-1}\}$ 
with the following relations

$$g_{i}g_{i+1}g_{i}=g_{i+1}g_{i}g_{i+1},$$

$$g_{i}g_{j}=g_{j}g_{i},~~\vert i-j\vert \geq 2,\eqno(1a)$$

$$e_{i}g_{i}=e_{i},$$

$$e_{i}g_{i-1}e_{i}=e_{i}.\eqno(1b)$$

\noindent Using these defining relations and via drawing pictures of link
diagrams [34-35], one can also derive other useful
ones. For example,

$$e_{i}e_{j}=e_{j}e_{i},~~~\vert i-j\vert\geq 2,$$

$$e_{i}^{2}=ne_{i},$$

$$(g_{i}-1)^{2}(g_{i}+1)=0.\eqno(1c)$$

\noindent It is easy to see that $\{ g_{1},~g_{2},~\cdots,~g_{f-1}\}$
generate a subalgebra ${\bf C}S_{f}$, which is isomorphic to the group
algebra of the symmetric group; that is, $D_{f}(n)\supset {\bf C}S_{f}$.
The properties of $D_{f}(n)$ have been discussed by many authors$^{[34-35]}$.
Based on these results, it is known that $D_{f}(n)$ is semisimple, i. e. it
is a direct sum of full matrix algebra over $\bf C$, when $n$ is not an integer
or is an integer with $n\geq f-1$, otherwise $D_{f}(n)$ is no longer semisimple.
In the following, we assume that the base field is ${\bf C}$ and $n$ is an integer
with $n\geq f-1$. In this case, $D_{f}(n)$ is semisimple, and irreducible
representations of $D_{f}(n)$ can be denoted by a Young diagram with
$f,~f-2,~f-4,~\cdots,~1$ or $0$ boxes. An irrep of $D_{f}(n)$ with
$f-2k$ boxes is denoted as $[\lambda ]_{f-2k}$. The branching rule
of $D_{f}(n)\downarrow D_{f-1}(n)$ is

$$[\lambda ]_{f-2k} =\oplus_{[\mu]\leftrightarrow [\lambda ]}[\mu ],\eqno(2)$$

\noindent where $[\mu ]$ runs through all the diagrams obtained by removing or
(if $[\lambda ]$ contains less than $f$ boxes) adding a box to $[\lambda ]$.
Hence, the basis vectors of $D_{f}(n)$ in the standard basis can be denoted
by

$$\left\vert
\matrix{
~~~~~[\lambda]_{f-2k} &D_{f}(n)\cr
[\mu ] &D_{f-1}(n)\cr
\vdots &\vdots\cr
[p] &D_{f-p+1}(n)\cr
[\nu ] &D_{f-p}(n)\cr}
\right) =
\left\vert\matrix{
[\lambda ]_{f-2k}\cr
[\mu ]\cr
\vdots\cr
[p]\cr
Y^{[\nu]}_{M}\cr}\right),\eqno(3)$$

\noindent where $[\nu ]$ is identical to the same irrep of $S_{f-p}$,
$Y^{[\nu ]}_{M}$
is a standard Young tableau, and $M$ can be understood either as the Yamanouchi
symbols or indices of the basis vectors in the so-called decreasing page order
of the Yamanouchi symbols. Procedures for evaluating matrix elements of
$g_{i}$, and $e_{i}$
with $i=1,~2,~\cdots,~f-1$ in the standard basis (3) have been given in
[36], and [37].
It is obvious that (3) is identical to the standard basis vectors of
$S_{f}$ when $k=0$.
In this case, all matrix elements of $e_{i}$ are zero, while matrix elements of
$g_{i}$ can be obtained by the well-known formula for $S_{f}$.
\vskip .5cm
\begin{centering}
{\large 3. Induced representations of $D_{f}(n)$}\\
\end{centering}
\vskip .4cm
   From the early work of Brauer [28] and recent studies [34-35] one knows
that there
is an important relation, the so-called Brauer-Schur-Weyl duality relation
between
the Brauer algebra $D_{f}(n)$ and O(n) or Sp(2m). If G is the orthogonal
group O(n) or
symplectic group Sp(2m), the corresponding centralizer algebra $B_{f}(G)$ are
quotients of Brauer's $D_{f}(n)$ or $D_{f}(-2m)$, respectively. We also need
a special class of Young diagrams, the so-called $n$-permissible Young diagrams
defined in [31]. A Young diagram $[\lambda ]$ is said to be $n$-permissible
if $P_{\mu}(n)\neq 0$ for all subdiagrams $[\mu ]\leq [\lambda ]$, where the
subdiagrams $[\mu ]$ can be obtained from $[\lambda ]$ by taking away
appropriate
boxes, and $P_{[\mu ]}(n)$ is the dimension of O(n) or $Sp(2m)$ for the irrep
$[\mu ]$. A Young diagram $[\lambda ]$ is $n$-permissible if and only if
\vskip .3cm
\noindent (a) Its first 2 columns contain at most $n$ boxes for $n$ positive,\\
\noindent (b) It contains at most $m$ columns for $n=-2m$ a negative even
integer,\\
\noindent (c) Its first 2 rows contain at most $2-n$ boxes for $n$ odd
and negative.
\vskip .3cm
\noindent If these conditions are satisfied, $D_{f}(n)$ is isomorphic to
$B_{f}(O(n))$ for $n$ positive, to $B_{f}(O(2-n))$ for $n$ negative and odd, and
to $B(Sp(2m))$ for $n=-2m<0$. In the following, we assume that all irreps to be
discussed are $n$-permissible with $n\leq f-1$ for $n>0$ or $-n\leq f-1$ for
negative $n$. These condition imply that the $D_{f}(n)$ being considered is
semisimple.
\vskip .3cm
   Therefore, an irrep of $B_{f}(O(n))$ or $B_{f}(Sp(2m))$ is simultaneously 
the same irrep of O(n) or Sp(2m). But the space of $B(G)$ and $G$
are different. The former is labelled by its Brauer algebra indices which
is operating among $B_{f}(G)$ space, while
the later
is labelled by its tensor components of group $G$. 
This is the so-called Brauer-Schur-Weyl
duality relation between $B_{f}(G)$ and $G$, where $G=O(n)$ or $Sp(2m)$.
\vskip .3cm
  Hence, in order to discuss the Kronecker products of $O(n)$ and $Sp(2m)$
for general
cases

$$[\lambda_{1}]\times [\lambda_{2}]\downarrow\sum_{\lambda}\{\lambda_{1}
\lambda_{2}\lambda\}[\lambda ],\eqno(4)$$

\noindent where $\{\lambda_{1}\lambda_{2}\lambda\}$ is the number of occurrence
of irrep $[\lambda ]$ in the decomposition $[\lambda_{1}]\times[\lambda_{2}]$,
we can switch to consider induced representations of the Brauer algebra,
$S_{f_{1}}\times S_{f_{2}}\uparrow D_{f}(n)$ for the same
decomposition given
by (4). In this case, we only need to study irreps of $D_{f}(n)$ induced by
irreps of
$S_{f_{1}}\times S_{f_{2}}$. The standard basis vectors of $[\lambda_{1}]_{f_{1}}$,
and $[\lambda_{2}]_{f_{2}}$ for $S_{f_{1}}$ and $S_{f_{2}}$ can be
denoted by $\vert Y^{[\lambda_{1}]}_{m_{1}}(\omega^{0}_{1})>$, and
$\vert Y^{[\lambda_{2}]}_{m_{2}}(\omega^{0}_{2})>$, respectively, where

$$(\omega^{0}_{1})=(1,~2,~\cdots,~f_{1}),~~~~(\omega_{2}^{0})
=(f_{1}+1,~f_{1}+2,~\cdots,~f_{1}+f_{2})\eqno(5)$$

\noindent are indices in the standard tableaux $Y^{[\lambda_{1}]}_{m_{1}}$
and $Y^{[\lambda_{2}]}_{m_{2}}$, respectively. The product of the two basis
vectors are denoted by

$$\vert Y^{[\lambda_{1}]}_{m_{1}},~Y^{[\lambda_{2}]}_{m_{2}},~(\omega_{1}^{0}),
~(\omega_{2}^{0})>\equiv \vert Y^{[\lambda_{1}]}_{m_{1}}(\omega_{1}^{0})>\vert
Y^{[\lambda_{2}]}_{m_{2}}(\omega^{0}_{2})>,\eqno(6)$$

\noindent which is called primitive uncoupled basis vector.$^{[31-32, 34]}$
\vskip .3cm
   The left coset decomposition of $D_{f}(n)$ with respect to the subalgebra
$S_{f_{1}}\times S_{f_{2}}$ is denoted by

$$D_{f}(n)=\sum_{\omega k}\oplus Q^{k}_{\omega}(S_{f_{1}}\times
S_{f_{2}}),\eqno(7)$$

\noindent where the left coset representatives $\{ Q^{k}_{\omega}\}$
have two types of operations. One is the order-preserving permutations,
which is the same as that for symmetric group,$^{[31-32]}$

$$Q^{k=0}_{\omega}(\omega^{0}_{1},~\omega^{0}_{2})=(\omega_{1},~\omega_{2}),
\eqno(8)$$

\noindent where

$$(\omega_{1})=(a_{1},~a_{2},~\cdots,~a_{f_{1}}),~~(\omega_{2})
=(a_{f_{1}+1},~a_{f_{1}+2},~\cdots,~a_{f})\eqno(9)$$

\noindent with $a_{1}<a_{2}<\cdots <a_{f_{1}}$,
$a_{f_{1}+1}<a_{f_{1}+2}<\cdots <a_{f}$, and $a_{i}$ represents any
one of the numbers $1,~2,~\cdots,~f$. The other, $\{ Q^{k\geq 1}_{\omega}\}$
contains $k$-time trace contractions between two sets of indices $(\omega_{1})$
and $(\omega_{2})$. For example, in $S_{2}\times S_{1}\uparrow D_{3}(n)$
for the outer product $[2]\times [1]$, there are six elements in
$\{ Q^{k}_{\omega}\}$ with

$$\{ Q^{0}_{\omega}\}=\{ 1,~g_{2},~g_{1}g_{2}\},~~\{Q^{1}_{\omega}\}
=\{ e_{2},~g_{1}e_{2},~e_{1}g_{2}\}.\eqno(10)$$

\noindent Let the number of operators in $\{ Q^{k}_{\omega}\}$ be $h$, and
the dimensions of the irreps $[\lambda_{1}]_{f_{1}}\times [\lambda_{2}]_{f_{2}}$
be $h_{[\lambda_{1}]}h_{[\lambda_{2}]}$, where $h_{[\lambda_{i}]}$ with $i=1,~2,$
can be computed, for example, by using Robinson's formula for symmetric group $S_{f}$. 
It is obvious that the total dimension
including multi-occurrence of the same irrep in the decomposition (4) is given
by $hh_{[\lambda_{1}]}h_{[\lambda_{2}]}$; that is,

$$hh_{[\lambda_{1}]}h_{[\lambda_{2}]}=
\sum_{\lambda}\{\lambda_{1}\lambda_{2}\lambda\}\dim ([\lambda
];~D_{f}(n)),\eqno(11)$$

\noindent where $\dim([\lambda ];D_{f}(n))$ is the dimension of $[\lambda
]$ for $D_{f}(n)$,
which was given in [29].
Hence, applying the $h$ $Q^{k}_{\omega}$'s to the primitive uncoupled
basis vector (6), we obtain all the uncoupled basis vectors needed in
construction of
the coupled basis vectors of $[\lambda ]$ for $D_{f}(n)$, which can be
denoted as

$$Q^{k}_{\omega}\vert Y^{[\lambda_{1}]}_{m_{1}},~Y_{m_{2}}^{[\lambda_{2}]},~(
\omega^{0}_{1}),~(\omega_{2}^{0})>=\vert
Y^{[\lambda_{1}]}_{m_{1}},~Y^{[\lambda_{2}]}_{m_{2}},~(\overbrace{\omega_{1}),~
(\omega_{2}}^{k})>,\eqno(12)$$

\noindent where $(\overbrace{\omega_{1}),~(\omega_{2}}^{k})$ stands for
$k$-time contractions between indices in $(\omega_{1})$ and $(\omega_{2})$.
However, all contractions among $(\omega_{1})$ or $(\omega_{2})$ will
be zero because $[\lambda_{i}]$ with $i=1,~2,$ has exact $f_{i}$ boxes, i.e.,
in this case, the irrep $[\lambda_{i}]$ of $S_{f_{i}}$ is the same irrep of 
$D_{f_{i}}(n)$. Therefore,  $S_{f_{1}}\times S_{f_{2}}$
can also be denoted as $D_{f_{1}}(n)\times D_{f_{2}}(n)$ when the irreps
$[\lambda_{i}]$ for $i=1,~2,$ has exactly $f_{i}$ boxes only. In the following,
we will always discuss this situation, and denote $S_{f_{1}}\times S_{f_{2}}$
as $D_{f_{1}}(n)\times D_{f_{2}}(n)$ without further explanation.
\vskip .3cm
   Finally, basis vectors of $[\lambda ]_{f-2k}$ can be expressed in terms
of the uncoupled
basis vectors given by (12).

$$\vert [\lambda ]_{f-2k},~\tau;~\rho >=
\sum_{m_{1}~m_{2}~\omega}C^{[\lambda]\rho;\tau}_{m_{1}m_{2};\omega}
Q^{k}_{\omega}\vert Y^{[\lambda_{1}]}_{m_{1}}(\omega_{1}^{0}),~Y^{[\lambda_{2}]}
_{m_{2}}(\omega^{0}_{2})>,\eqno(13)$$

\noindent where $\rho$ is the multiplicity label needed in the outer-product
$[\lambda_{1}]_{f_{1}}\times [\lambda_{2}]_{f_{2}}\uparrow[\lambda]_{f-2k}$,
$\tau$ stands for other labels needed for the irrep $[\lambda ]_{f-2k}$, and
the coefficient $C^{[\lambda ]\rho;\tau}_{m_{1}m_{2};\omega}$ is
$[\lambda_{1}]_{f_{1}}\times[\lambda_{2}]_{f_{2}}\uparrow[\lambda]_{f-2k}$
induction coefficient (IDC) or the outer-product reduction coefficient (ORC).
\vskip .5cm
\begin{centering}
{\large 4. A Young diagrammatic method for Kronecker products\\
 of O(n) and Sp(2m)}\\
\end{centering}
\vskip .3cm
   Analytical derivation or algorithm for the IDCs discussed in Section 3
is not necessary if only outer-products of $D_{f_{1}}(n)\times
D_{f_{2}}(n)$ for irreps $[\lambda_{1}]_{f_{1}}\times [\lambda_{2}]_{f_{2}}$
are
considered. It is obvious in (12) that irreps with $f-2k$ boxes of
$D_{f}(n)$ can
be induced from irreps of $D_{f_{1}}(n)\times D_{f_{2}}(n)$. When $k=0$, (12) is
identical to that for symmetric groups. An important operation in (12)
is performed by $\{ Q^{k}_{\omega}\}$ with $k\neq 0$. After $k$-time
contraction the
uncoupled primitive basis vector of $[\lambda_{1}]_{f_{1}}\times [\lambda_{2}]
_{f_{2}}$ will be equivalent to basis vectors of
$[\lambda_{1}^{\prime}]_{f_{1}-k}\times [\lambda_{2}^{\prime}]_{f_{2}-k}$,
where $[\lambda^{\prime}_{i}]_{f_{i}-k}$ with $i=1,~2$ is any possible
standard Young
diagrams with $f_{i}-k$ boxes, which can be obtained from
$[\lambda_{i}]_{f_{i}}$
by deleting $k$ boxes from $[\lambda_{i}]$ in all possible ways.
Therefore, as far as representations are concerned,
the irrep $\{[\lambda]_{f-2k}\}$
of $D_{f}(n)$ can be obtained from the outer-product $\{
[\lambda^{\prime}_{1}]_{f_{1}-k}
\times[\lambda^{\prime}_{2}]_{f_{2}-k}\}$ of the symmetric group
$S_{f_{1}-k}\times S_{f_{2}-k}$.
Thus, we obtain the following rules for the outer-products of
$D_{f_{1}}(n)\times D_{f_{2}}(n)$.
\vskip .5cm
\noindent {\bf Lemma 1.} The outer-product rule for
$D_{f_{1}}(n)\times D_{f_{2}}(n)\uparrow D_{f}(n)$ for the decomposition

$$[\lambda_{1}]_{f_{1}}\times [\lambda_{2}]_{f_{2}}\uparrow
\sum_{\lambda}\{\lambda_{1}\lambda_{2}\lambda\}[\lambda]$$

\noindent can be obtained diagrammatically by
\vskip .3cm
\noindent (1) Removing $k$ boxes, where $k=0,~1,~2,~\cdots,\min(f_{1},f_{2})$,
from $[\lambda_{1}]_{f_{1}}$ and $[\lambda_{2}]_{f_{2}}$ simultaneously in
all possible ways under the following restrictions:
\vskip .3cm
\noindent (a) Always keep the resultant diagrams
$[\lambda_{i}^{\prime}]_{f_{i}-k}$
with $i=1,~2$ standard Young diagrams;
\vskip .3cm
\noindent (b) No more than two boxes in the same column (row) in $[\lambda_{1}]$
with those in the same row (column) in $[\lambda_{2}]$ can be removed
simultaneously.
\vskip .3cm
\noindent (2) Applying the Littlewood rule of the outer-product reduction for
symmetric group to the outer-product
$[\lambda^{\prime}_{1}]_{f_{1}-k}\times [\lambda_{2}^{\prime}]_{f_{2}-k}$, and
repeatedly doing so for each $k$.
\vskip .4cm
   What we need to explain is restriction (b). Consider a simple
example which is representative of the general case. Let $[\lambda_{1}]=[2]$,
$[\lambda_{2}]=[1^{2}]$, and  a $k$-time trace contraction operator be $Q^{k}$.
According to our procedure, we have

$$Q^{1}(~~
\begin{tabular}{|l|l|}
\hline
{~} &{~~}\\
\hline
\end{tabular}~\times~~
\begin{tabular}{|l|}
\hline
{~}\\
\hline
{~~}\\
\hline
\end{tabular}~~)=(~~
\begin{tabular}{|l|l|}
\hline
{~} &$\alpha$\\
\hline
\end{tabular}~\times~
\begin{tabular}{|l|}
\hline
{~}\\
\hline
$\alpha$\\
\hline
\end{tabular}~~)=(~
\begin{tabular}{|l|}
\hline
{~}\\
\hline
\end{tabular}~\times~
\begin{tabular}{|l|}
\hline
{~}\\
\hline
\end{tabular}~),\eqno(14a)$$

\noindent while

$$Q^{2}(~~
\begin{tabular}{|l|l|}
\hline
{~} &{~~}\\
\hline
\end{tabular}~\times~
\begin{tabular}{|l|}
\hline
{~}\\
\hline
{~~}\\
\hline
\end{tabular}~)=(~
\begin{tabular}{|l|l|}
\hline
$\beta$ &$\alpha$\\
\hline
\end{tabular}~\times~
\begin{tabular}{|l|}
\hline
$\beta$\\
\hline
$\alpha$\\
\hline
\end{tabular}~).\eqno(14b)$$
\vskip .3cm
\noindent Because trace contraction occurs in pairs, the indices
$\alpha$, and $\beta$ labelled in the boxes indicate that those
with the same indices are contracted with each other. It is known
that trace contraction of two vectors results in the symmetrization
of the tensor components. Therefore, trace contraction of anti-symmetric
tensors is zero. However, the indices of $\alpha$ part is not only symmetric
but also anti-symmetric with those of $\beta$ part in (14b). Hence,
restriction (b)
holds.
\vskip .3cm
   Finally, using the Brauer-Schur-Weyl duality relation between
$D_{f}(n)$ and O(n) or Sp(2m), one knows that Lemma 1 applies
to the decompositions of the Kronecker products of O(n) or Sp(2m)
as well. Thus, we have the following lemma.
\vskip .3cm
\noindent {\bf Lemma 2.} The Kronecker product of O(n) or Sp(2m)
for the decomposition given by (4) can be obtained by using procedures
(1) and (2) given by Lemma 1 together with the following modification
rules:
\vskip .3cm
\noindent  For the group O(n), where $n=2l~{\rm or}~2l+1$, (Sp(n), where
$n=2l$), the resulting
irrep $[\lambda ]=[\lambda_{1},~\lambda_{2},~\cdots,\lambda_{p},~\dot{0}]$
is nonstandard if $p>l$. In this case, we need to remove boxes from
$[\lambda ]$ along a continuous boundary with hook of length $2p-n$ ($2p-n-2$)
and depth $x$, where $x$ is counted by starting from the first column of
$[\lambda ]$ to
the right-most column that the boundary hook reaches.$^{[12]}$  The
resultant Young diagram
will be admissible or set to zero if, at any stage, the removal of the
required hook
leaves an irregular Young diagram. Then, the resultant irrep $[\lambda
]_{\rm allowed }$
can be denoted  symbolically as

$$[\lambda ] _{\rm allowed}\left\{
\begin{array}{l}
=(-)^{x}[\sigma ],~~~{\rm for~~O(n)},\\
{}\\
=(-)^{x+1}[\sigma ],~~~{\rm for~~Sp(2m),}
\end{array}\right.$$

\noindent where $[\sigma ]$ is obtained from $[\lambda ]$ by using
the above modification rules. For example,

$$[3^{3},~1]=\left\{
\begin{array}{l}
=[3^{3}]~~~{\rm for~ O(7)},\\
\\
=[3^{2}]~~~{\rm for~O(4)},\\
\\
=-[20]~~~{\rm for~ O(2)},\\
\\
=0~~~~~{\rm for~O(6),~O(5),~and~O(3),}
\end{array}\right.\eqno(15)$$

\noindent which was illustrated in [12] by King. In the following, we give
an example to show how this method works.
\vskip .4cm
\noindent {\bf Example.} Find the Kronecker product $[21]\times [11]$ for
O(n) or Sp(2m).
\vskip .3cm
\noindent  First, we consider all possible diagrams with 0, 1, and
$\min(f_{1},~f_{2})=2$ -time trace contractions, which are

$$\begin{array}{l}
{\begin{tabular}{|l|l|}
\hline
{~} &{~~}\\
\hline
\end{tabular}}\\
{\begin{tabular}{|l|}
{~~}\\
\hline
\end{tabular}}
\end{array}~\times~
\begin{tabular}{|l|}
\hline
{~~}\\
\hline
{~~}\\
\hline
\end{tabular}~,~~
\begin{array}{l}
{\begin{tabular}{|l|l|}
\hline
{~} &{$\tiny\times$}\\
\hline
\end{tabular}}\\
{\begin{tabular}{|l|}
\hline
{~~}\\
\hline
\end{tabular}}
\end{array}~\times~
\begin{tabular}{|l|}
\hline
{~}\\
\hline
{$\tiny\times$}\\
\hline
\end{tabular}~,~
\begin{array}{l}
{\begin{tabular}{|l|l|}
\hline
{~~} &{~~}\\
\hline
\end{tabular}}\\
{\begin{tabular}{|l|}
{$\tiny\times$}\\
\hline
\end{tabular}}
\end{array}~\times~
\begin{tabular}{|l|}
\hline
{~~}\\
\hline
{$\tiny\times$}\\
\hline
\end{tabular}~,~~
\begin{array}{l}
{\begin{tabular}{|l|l|}
\hline
{~~} &{$\tiny\times$}\\
\hline
\end{tabular}}\\
{\begin{tabular}{|l|}
\hline
{$\tiny\times$}\\
\hline
\end{tabular}}
\end{array}~\times~
\begin{tabular}{|l|}
\hline
{$\tiny\times$}\\
\hline
{$\tiny\times$}\\
\hline
\end{tabular}~.\eqno(16)$$

\noindent Then, we need to compute the Kronecker products
$[21]\times [11]$, $[11]\times [1]$, $[2]\times [1]$, and $[1]
\times [0]$, which can be obtained by using the Littlewood rule for U(n). We get

$$[21]\times [11]=[32]+[221]+[2111]+[311],\eqno(17a)$$

$$[20]\times [1]=[30]+[21],\eqno(17b)$$

$$[11]\times[1]=[21]+[111],\eqno(17c)$$

$$[1]\times [0]=[1].\eqno(17d)$$

\noindent Finally, summing up all the irreps appearing on the
rhs. of (17), we obtain

$$[21]\times [11]=[32]+[221]+[2111]+[311]+[30]+2[21]+[111]
+[10],\eqno(18)$$

\noindent which is valid for O(n) when $n\geq 8$
and Sp(2m) when $m\geq 4$. Using the modification rules
given in Lemma 2, we can easily obtain the following
results

$$[210]\times [110]=[320]+[221]+[211]+[311]+[300]+2[210]+[111]+[100]~~
{\rm for~ O(7)},\eqno(19a)$$

$$[210]\times [110]=[320]+[221]+3[210]+[311]+[300]+[111]+[100]~~
{\rm for~ O(6)},\eqno(19b)$$

$$[21]\times [11]=[32]+[22]+[20]+[31]+[30]+2[21]+[11]+[10]~~
{\rm for~ O(5)},\eqno(19c)$$

$$[21]\times [11]=[32]+2[30]+2[21]+2[10]~~{\rm for~ O(4)}.\eqno(19d)$$

\noindent In the above computation, the following results have been used

$$[2111]=\left\{
\begin{array}{l}
{[211]~~{\rm for~~O(7)}},\\
{[21]~~{\rm for~~O(6)}},\\
{[20]~~{\rm for~~O(5)}},\\
{[10]~~{\rm for ~~O(4)}},
\end{array}\right.~~~\eqno(20a)$$

$$[221]=\left\{
\begin{array}{l}
{[22]~~{\rm for~~O(5)}},\\
{0~~~~~~{\rm for~~O(4)}},\\
\end{array}\right.\eqno(20b)$$

$$[311]=\left\{
\begin{array}{l}
{[31]~~{\rm for~~ O(5)}},\\
{[30] ~~{\rm for~~ O(4)}},
\end{array}\right.\eqno(20c)$$

\noindent which are obtained from modification rules given in Lemma 2.
While

$$[210]\times [110]=[320]+[221]+[311]+[300]+2[210]+[111]+
[100]~~{\rm for ~~Sp(6)},~~~~~~~~\eqno(21a)$$

$$[21]\times [11]=[32]+[30]+[21]+[10]~~{\rm for ~~Sp(4)},\eqno(20b)$$

\noindent where the following modification rule have been used:

$$[2111]=\left\{
\begin{array}{l}
{~~~0~~{\rm for~~~~~Sp(6)}},\\
{-[21]~~{\rm for~~Sp(4)}}
\end{array}\right.,\eqno(22a)$$

$$[221]=[311]=[111]=0~~{\rm for~Sp(4)}.\eqno(22b)$$
\vskip .5cm
\begin{centering}
{\bf 5. Concluding Remarks}\\
\end{centering}
\vskip .3cm
   In this paper, a new simple Young diagrammatic method for the decomposition
of the Kronecker products of O(n) and Sp(2m) is outlined based on the induced
representation theory of $D_{f}(n)$. This algebra was proposed by Brauer at the
end of thirties. His aim was indeed to solve the decomposition problem of the
Kronecker products of O(n) and Sp(2m). On the other hand, because the
representations
of $D_{f}(n)$ are the same as those of Birman-Wenzl algebras $C_{f}(r,q)$
when  $r$,
and $q$ are not a root of unity, the method applies to quantum groups
$O_{q}(n)$ and
$Sp_{q}(2m)$ as well for $q$ being not a root of unity. The induced
representations
of $D_{f}(n)$ presented in Section 3 can also be used to derive Clebsch-Gordan
coefficients of SO(n) when IDCs of $D_{f_{1}}(n)\times D_{f_{2}}(n)$ are
evaluated,
which will be discussed in our next paper.
\vskip .3cm
   It should be stated that though our Young diagrammatic method for decomposition
of $O(n)$ and $Sp(2m)$ Kronecker products is derived from induced representation
theory of Brauer algebra with the help of Brauer-Schur-Weyl duality relation,
the final results are the same as those derived  by Littlewood and Newell
based on character theory and Schur functions$^{[18-19]}$. In[18], the main results
on how to obtain the Kronecker product of $O(n)$ and $Sp(2m)$ were achieved through
the combinatorials of certain type of S-functions. However, in [18], only $p\geq r$
cases were considered, where $n=2p$ or $2p+1$ for $O(n)$, and $p=m$ for $Sp(2m)$, and
$r$ is the number of rows for the corresponding irrep. In this case, no modification
rule is needed, which is the same as ours. When $p\leq r$ in a Young diagram, the
final diagram with number of rows greater than $p$ will become non-standard irrep,
the correspondence between these non-standard diagrams and the corresponding
standard ones with signs in the front of the diagrams was first studied by Newell
in [19], which gives just the so-called modification rules proposed by King
in a much simper manner.$^{[12]}$ This fact is now summarized by Lemma 2 in this paper.
\vskip .3cm
On the other hand, the Young tableau method proposed by
Littelmann [24] and crystal graph base given in [25] are related to the weight space
of the corresponding Lie groups (algebras). Therefore, these methods  do not use the
representation theory of symmetric groups at all. But the final results on
the decomposition of the Kronecker product of $O(n)$ and $Sp(2m)$ are the same
as those obtained by our Young diagrammatic method derived from Brauer algebras.
\vskip .3cm
   Furthermore, this method can also be applied to the Kronecker products of
SO$(2l+1)$ for any irreps and SO$(2l)$ for their irreps
$[\lambda_{1},~\lambda_{2},~\cdots,\lambda_{k},\dot{0}]$
for $k<l$. If $k=l$, the irrep of O(2l)
$[\lambda_{1},~\lambda_{2},~\cdots,~\lambda_{k}]$ with $\lambda_{k}\neq 0$
reduces into irreps of SO(2l) denoted by $[\lambda_{1},~\lambda_{2},~\cdots,~
\lambda_{k}]$ and $[\lambda_{1},~\lambda_{2},~\cdots,~-\lambda_{k}]$, of which
the dimensions are the same.
In this case, one
should be cautious and use this method. The dimension formula for SO(n) is
always helpful
in checking final results.
\vskip .3cm
   Finally, it should be noted that the method applies to tensor or
``true'' representations
of O(n) only. The spinor representations of O(n) are related to spinor representations of
Brauer algebras according to the Brauer-Schur-Weyl duality relation, which
still need to be
further studied.

\vskip .8cm
\noindent {\bf Acknowledgment}
\vskip .5cm
  The authors are very much grateful to our referees for their helpful
suggestions and comments, especially for providing us with references [18-19],
and [24-26] which we formerly let unnoticed.
 The project was supported by  National Natural Science Foundation
of China, and a grant from US National Science Foundation
through LSU.
\vskip .5cm
\begin{tabbing}
\=1111\=22222222222222222222222222222222222222222222222222222222222222222222
222222222222\=\kill\\
\>{[1]}\>{B. G. Wybourne, Int. J. Quant. Chem., {\bf 7}(1973) 117}\\
\>{[2]}\>{H. Fukutome, M. Yamamura, and A. Nishiyama, Prog. Theor. Phys.,
{\bf 57} (1977) 1554}\\
\>{[3]}\>{M. Gell-Mann, Rev. Mod. Phys., {\bf 50}(1978) 721}\\
\>{[4]}\>{M. Gell-Mann, P. Ramond, and R. Slansky, Nucl. Phys., {\bf
B159}(1979) 141}\\
\>{[5]}\>{A. Arima, and F. Iachello, Ann. Phys. (N. Y.), {\bf 99}(1976) 253}\\
\>{[6]}\>{C. L. Wu, Da Hsuan Feng, M. Guidry, Adv. Nucl. Phys. {\bf 21}
(1994) 227}\\
\>{[7]}\>{Feng Pan, Zeng-Yong Pan, and Yu-Fang Cao, Chin. Phys. Lett., {\bf
8} (1991) 56}\\
\>{[8]}\>{P. Navratil, H. B. Geyer, and J. Dobaczewski, Ann. Phys. (N. Y.),
{\bf 243}
(1995) 218}\\
\>{[9]}\>{G. Rosensteel, and D. J. Rowe, Phys. Rev. Lett., {\bf 38} (1977) 10}\\
\>{[10]}\>{O. Casta\~{n}os, P. Hess, J. P. Draayer, and P. Pochford, Nucl.
Phys. {\bf A524} (1991) 469}\\
\>{[11]}\>{R. C. King and N. G. I. El-Sharkaway, J. Phys. {\bf A16} (1983)
3157}\\
\>{[12]}\>{R. C. King, J. Math. Phys. {\bf 12}(1971) 1588}\\
\>{[13]}\>{R. C. King, Lect. Notes in Phys.  {\bf 50}(1975) 481 }\\
\>{[14]}\>{R. C. King, Luan Dehuai, and B. G. Wybourne, J. Phys.  {\bf A14}
(1981) 2509}\\
\>{[15]}\>{R. C. King, J. Phys. {\bf A8}(1975) 429}\\
\>{[16]}\>{F. D. Murnaghan, The Theory of Gourp Representations, (Johns
Hopkings, Baltimore,1938)}\\
\>{[17]}\>{D. E. Littlewood, The Theory of Gourp Characters, 2nd edn.
(Oxford, Claredom, 1950)}\\
\>{[18]}\>{D. E. Littlewood, Can. J. Math. {\bf 10}(1958) 17}\\
\>{[19]}\>{M. J. Newell, Proc. Royal Irish Acad. {\bf 54A}(1951) 153}\\
\>{[20]}\>{K. Koike, and I. Terada, J. Alg., {\bf 107}(1987) 466}\\
\>{[21]}\>{M. Fischer, J. Math. Phys. {\bf 22} (1981) 637}\\
\>{[22]}\>{G. Girardi, A. Sciarrino, and P. Sorba, J. Phys. {\bf A15}
(1982) 1119}\\
\>{[23]}\>{G. Girardi, A. Sciarrino, and P. Sorba, J. Phys.  {\bf A16}
(1983) 2069}\\
\>{[24]}\>{P. Littelmann, J. Alg. {\bf 130}(1990) 328}\\
\>{[25]}\>{T. Nakashima, Commun. Math. Phys. {\bf 154}(1993) 215}\\
\>{[26]}\>{M. Kashiwara, J. Alg. {\bf 165}(1994)295}\\
\>{[27]}\>{H. Weyl, The Classical Groups, (Princeton  U. P., Princeton, N.
J., 1939)}\\
\>{[28]}\>{R. Brauer, Ann. Math., {\bf 63} (1937) 854}\\
\>{[29]}\>{A. Berele, J. Comb. Th. Series A, {\bf 43} (1986) 320}\\
\>{[30]}\>{S. Sundaram, J. Comb. Th. Series A, {\bf 53} (1990) 239}\\
\>{[31]}\>{Feng Pan, and J. Q. Chen, J. Math. Phys. {\bf 34} (1993) 4305;
4316}\\
\>{[32]}\>{Feng Pan, J. Phys. {\bf A26}(1993) 4621}\\
\>{[33]}\>{Feng Pan, and Lianrong Dai, J. Phys., {\bf A29} (1996) 5079; 5093}\\
\>{[34]}\>{H. Wenzl, Ann. Math., {\bf 128}(1988) 173}\\
\>{[35]}\>{H. Wenzl, Commun. Math. Phys., {\bf 313}(1990) 383}\\
\>{[36]}\>{Feng Pan, J. Phys., {\bf A28} (1995) 3139}\\
\>{[37]}\>{R. Leduc, A. Ram, 1996, to appear in Adv. Math.}\\

\end{tabbing}
\end{document}